\documentclass[english]{article}
\usepackage[T1]{fontenc}
\usepackage[latin9]{inputenc}
\usepackage{amsmath}
\usepackage{amssymb}
\usepackage{esint}

\makeatletter

\usepackage{babel}

\makeatother

\usepackage{babel}
\begin{document}
\begin{titlepage} \thispagestyle{empty}

\bigskip{}

\noindent \begin{center}
{\Large {Quantum Scalar Field in D-dimensional de Sitter Spacetimes}}\\

\par\end{center}

\begin{center}
\vspace{0.5cm}

\par\end{center}

\noindent \begin{center}
{G. Alencar $^{a}$%
\footnote{e-mail: geovamaciel@gmail.com %
}, I. Guedes $^{b}$, R. R. Landim $^{b}$ and R.N. Costa Filho $^{b}$}
\par\end{center}

\begin{center}
\vspace{0.5cm}
 \textit{$^{a}$Universidade Estadual do Cear\'a, Faculdade de Educa\c c\~ao, Ci\^encias e Letras do Sert\~ao Central-R. 
Epitácio Pessoa, 2554, 63.900-000  Quixad\'{a}, Cear\'{a},  Brazil. \vspace{0.2cm}
 } \textit{$^{b}$Departamento de F\'{\i}sica, Universidade Federal do Cear\'{a},
Caixa Postal 6030, Campus do Pici, 60455-760, Fortaleza, Cear\'{a}, Brazil. }
\par\end{center}

\vspace{0.3cm}

\begin{abstract}
In this work we investigate the quantum theory of scalar fields propagating
in a $D-$dimensional de Sitter spacetime. The method of dynamic invariants
is used to obtain the solution of the time-dependent Schrödinger equation.
The quantum behavior of the scalar field in this background is analyzed,
and the results generalize previous ones found in the literature. We point that
the Bunch-Davies thermal bath depends on the choice of $D$ and the conformal parameter $\xi$.
This is important in extra dimension physics, as in the Randall-Sundrum
model. 
\end{abstract}
\end{titlepage}

\section{Introduction}

Although String Theory is a promising solution to the quantization of gravity,\cite{Polchinski:1998rq,Polchinski:1998rr,Berkovits:2000fe},
in cosmological scales gravity field can be considered as a classical theory
and the fields as propagating waves in the background. Gauge and scalar
fields, for instance, can be quantized in this background by the use
of semiclassical approach \cite{Birrell:1982ix}. This is very similar
to the early days of Quantum Field Theory. The quantization of matter
was very established but the quantization of the fields was not well
understood. Therefore, many calculations were done considering fields
as backgrounds. Time-dependent backgrounds are used to describe many
physical systems yielding interesting results. For instance, in the
study of black hole evaporation \cite{Hawking:1974rv}, the Unruh,
and Casimir effects \cite{Crispino:2007eb,Saharian:2009ii}. They
are also very useful to describe the dynamical evolution of the universe,
where the production of particles in cosmological spacetimes has been
investigated \cite{Parker:1968mv,Parker:1969au,Parker:1971pt}.

The core idea of extra dimensional models is to consider the four-dimensional
universe as a hyper-surface embedded in a multidimensional manifold.
After the proposal of Kaluza and Klein, this idea attracted not much
attention. This changed a lot after the advent of supergravity and
superstring theory, where extra dimensions are a necessary ingredient.
More recently, after the Randall and Sundrum proposal of a Brane world
with non factorisable metric there have been an extensive use of these
ideas \cite{Randall:1999vf,Randall:1999ee}. This model provides a
possible solution to the hierarchy problem and show how gravity is
trapped to a membrane. After while, this model has been modified to
consider the membrane as a topological defect generated by a scalar
field \cite{Kehagias:2000au,Bazeia:2008zx}. This solves some problems
related to the localization of fields in the membrane \cite{Landim:2010pq,Landim:2011ki}.

In 2004, Carvalho, Furtado, and Pedrosa \cite{MM} investigated the
quantum scalar fields in a Friedman-Robertson-Walker (FRW) background.
They demonstrated that the problem of the field quantization in this
background reduces to solve the time-dependent Schrödinger equation
for the harmonic oscillator with time-dependent mass and frequency.
To solve the time-dependent Schrödinger equation (TDSE), they employed
the dynamical method of Lewis and Riesenfeld \cite{Lewis:1968tm}.
By considering a quadratic invariant (I), they found the exact solution
of the problem and established the existence of squeezed states in
this background.

The quantum effects of a massive scalar field in the de Sitter spacetime
was investigated in Ref.\cite{pedrosa1}, where Lopes \textit{et al.}
used exact linear invariants and the Lewis and Riesenfeld method to
derive the corresponding Schrödinger states in terms of solutions
of a second-order ordinary differential equation. They also constructed
Gaussian wave packet states and calculate the quantum dispersion as
well as the quantum correlations for each mode of the quantized scalar
field. Aspects related to Bunch-Davies vacuum for the scalar field
using the same method has been analyzed in \cite{Bertoni:1997qb}.

The calculations presented in Refs. \cite{pedrosa1,Bertoni:1997qb}
were performed in a $D=4$ spacetime. Here we intend to generalize
the quantization of the scalar field in a D-dimensional spacetime
also by using the Lewis and Riesenfeld method. First, we analyze the
scalar quantum field in a D-dimensional FRW background, and, second
we obtain the exact solution for a $D-$dimensional de Sitter spacetime.

\section{Decomposition of the Scalar Field}
Consider the scalar field in a $D-$dimensional Friedmann-Robertson-Walker
(FRW) spacetime. The Lagrangian density for the scalar field is given
by

\begin{equation}
{\cal L}=-\frac{1}{2}g^{\mu\nu}\partial_{\mu}\Phi\partial\Phi-\frac{1}{2}\xi R\Phi^{2}-\frac{1}{2}\mu\Phi^{2},
\end{equation}
where the metric $ds^{2}=-dt^{2}+a^{2}(t)d\vec{x}\cdot d\vec{x}$ and $R$ is the
Ricci scalar. The scalar field can be decomposed in a complete basis
$u_{k}$ given by 
\begin{equation}
u_{k}(\vec{x},t)=e^{i\vec{k}\cdot\vec{x}}\phi_{k}(t)\equiv e^{i\vec{k}\cdot\vec{x}}\frac{\phi_{k}^{1}(t)+i\phi_{k}^{2}(t)}{\sqrt{2}},
\end{equation}
 where $i=1,2$ labels the real and imaginary parts of $\phi_{k}$. With
the definition $\omega^{2}=k^{2}/a^{2}+\mu^{2}+\xi R$ the action
reads 
\begin{equation}
S=\frac{1}{2}\sum_{i=1,2}\int dt \int \frac{d^{D-1}k}{(2\pi)^{D-1}}a^{(D-1)}\left[\dot{\phi}_{k}^{i}{}^{2}-\omega^{2}(t)\phi_{k}^{i}{}^{2}\right].
\end{equation}
From the above action we obtain the Hamiltonian for the each mode
of the scalar field 
\begin{equation}
H_{ik}=\frac{1}{2}(a^{-(D-1)}(t)p_{ik}^{2}+\omega^{2}a^{(D-1)}(t)q_{ik}^{2}),
\end{equation}
 where 
\begin{equation}
p_{ik}=\frac{\partial L}{\partial\dot{\phi}_{ik}}=a(t)^{(D-1)}\dot{\phi}_{ik},
\end{equation}
 with $p$ being the conjugate momentum. The classical equation of
motion for the $qth$ mode reads 
\begin{equation}
\ddot{q}_{\lambda ik}+(D-1)\frac{\dot{a}}{a}\dot{q}_{ik}+\omega^{2}q_{ik}^{2}=0.\label{classicaleq}
\end{equation}
Next, consider the classical harmonic oscillator with time-dependent
mass and frequency given by the Hamiltonian 
\begin{equation}
H(t)=\frac{p^{2}}{m(t)}+\frac{1}{2}m(t)\omega^{2}(t)q^{2},\label{hamiltoniano}
\end{equation}
where $[q,p]=i\hbar$. The equation of motion reads 
\begin{equation}
\ddot{q}+\frac{\dot{m}(t)}{m(t)}\dot{q}+\omega^{2}(t)q=0,\label{HO}
\end{equation}
which is similar to Eq. (\ref{classicaleq}) if one considers that
each mode of the scalar field corresponds to the time-dependent
harmonic oscillator with $m(t)=a^{D-1}(t)$ and $\omega=(k^{2}/a^{2}+\mu^{2}+\xi R)^{\frac{1}{2}}$.

\section{Quantization of the Scalar Field with Emarkov Approach}
Consider a time-dependent harmonic oscillator described by Eq. (\ref{hamiltoniano}).
It is well known that an invariant for Eq. (\ref{hamiltoniano}) is
given by \cite{Carinena}
\begin{equation}
I=\frac{1}{2}\left[\left(\frac{q}{\rho}\right)^{2}+(\rho p-m\dot{\rho q})^{2}\right]\label{invariantedef}
\end{equation}
where $q(t)$ satisfies Eq.(\ref{HO}) and $\rho(t)$ satisfies the
generalized Milne-Pinney equation \cite{Milne,Pinney}

\begin{equation}
\ddot{\rho}+\gamma(t)\dot{\rho}+\omega^{2}(t)\rho=\frac{1}{m^{2}(t)\rho^{3}}\label{MP}
\end{equation}
with $\gamma (t)=\dot{m}(t)/m(t)$. The invariant $I(t)$ satisfies the equation

\begin{equation}
\frac{dI}{dt}=\frac{\partial I}{\partial t}+\frac{1}{i\hbar}[I,H]=0
\end{equation}
and can be considered hermitian if we choose only the real solutions
of Eq. (\ref{MP}). Its eingenfunctions, $\phi_{n}(q,t)$, are assumed
to form a complete orthonormal set with time-independent discrete eigenvalues,
$\lambda_{n}=(n+\frac{1}{2})\hbar$. Thus
\begin{equation}
I\phi_{n}(q,t)=\lambda_{n}\phi_{n}(q,t)\label{invariante}
\end{equation}
with $\left\langle \phi_{n},\phi_{n'}\right\rangle =\delta_{nn'}$.
Taking he Schr\"odinger equation (SE)

\begin{equation}
i\hbar\frac{\partial\psi(q,t)}{\partial t}=H(t)\psi(q,t)\label{SE}
\end{equation}
where $H(t)$ is given by Eq. (\ref{hamiltoniano}) with $p=-i\hbar\frac{\partial}{\partial q}$,
Lewis and Riesenfeld \cite{Lewis:1968tm} showed that the solution $\psi_{n}(q,t)$
of the SE (see Eq.(\ref{SE})) is related to the functions $\phi_{n}(q,t)$
by 

\begin{equation}
\psi_{n}(q,t)=e^{i\theta_{n}(t)}\phi_{n}(q,t)
\end{equation}
where the phase functions $\theta_{n}(t)$ satisfy the equation

\begin{equation}
\hbar\frac{d\theta_{n}(t)}{dt}=\left\langle \phi_{n}(q,t)\left|i\hbar\frac{\partial}{\partial t}-H(t)\right|\phi_{n}(q,t)\right\rangle .
\end{equation}
The general solution of the SE may be written as

\begin{equation}
\psi (q,t)=\sum_{n}c_{n}e^{i\theta_{n}(t)}\phi_{n}(q,t)
\end{equation}
where $c_{n}$ are time-independent coefficients. Now, using an unitary transformation and following the steps drawn in Ref. \cite{Carinena} we find

\begin{align}\label{psi}
\psi_{n}(q,t)=&e^{i\theta_{n}(t)}\left(\frac{1}{\pi^{1/2}\hbar^{1/2}n!2^{n}\rho}\right)^{1/2}\times\nonumber\\
&\exp\left\lbrace\frac{im(t)}{2\hbar}\left[\frac{\dot{\rho}}{2\hbar}+\frac{i}{m(t)\rho^{2}(t)}\right]q^{2}\right\rbrace\times \\
&H_{n}\left(\frac{1}{\sqrt{\hbar}}\frac{q}{\rho}\right)\nonumber
\end{align}
where 

\begin{equation}
\theta_{n}(t)=-(n+\frac{1}{2})\int_{t_{0}}^{t}\frac{1}{m(t')\rho^{2}}dt'\label{theta},
\end{equation}
and $H_{n}$ is the Hermite
polynomial of order $n$. Using the mass and frequency defined previously Eq. (\ref{MP}) reads 
\begin{equation}
\ddot{\rho}+(D-1)\frac{\dot{a}}{a}\dot{\rho}+\left[\frac{k^{2}}{a^{2}}+\mu^{2}+\xi R\right]\rho=\frac{a^{-2(D-3)}(t)}{\rho^{3}}.\label{MPa}
\end{equation}

To find the exact solutions of Eq. (\ref{psi}), one has to solve
Eq.(\ref{MPa}) or find the two linearly independent solutions of
Eq. (\ref{classicaleq}). Let us consider the latter case. Let $dt=a(t)d\eta$
be the conformal time and let us define a new variable $q_{\lambda ik}=\Omega\bar{q}_{\lambda ik}$.
With this, Eq. (\ref{classicaleq}) reads
\begin{eqnarray}
 &  & \bar{q}''+\left[2a\frac{\dot{\Omega}}{\Omega}-\dot{a}+(D-1)\dot{a}\right]\bar{q}_{ik}'+\nonumber \\
 &  & [(k^{2}+a^{2}\mu^{2}+a^{2}\xi R)+a^{2}\frac{\ddot{\Omega}}{\Omega}+(D-1)a\dot{a}\frac{\dot{\Omega}}{\Omega}]\bar{q}_{ik}=0\label{classicaleqD}
\end{eqnarray}
where the prime and the dot means a derivative with respect to $\eta$
and $t$ respectively. By choosing $\Omega=a^{-(D-1)/2}$ one finds
\begin{eqnarray}\label{qbarra}
 &  & \bar{q}_{\lambda ik}''-\dot{a}\bar{q}_{\lambda ik}'+[(k^{2}+a^{2}\mu^{2}+a^{2}\xi R)+\frac{(D-1)(D+1)}{4}\dot{a}^{2}\nonumber \\
 &  & -\frac{(D-1)}{2}a\ddot{a}-\frac{(D-1)^{2}}{2}\dot{a}^{2}]\bar{q}_{\lambda ik}=0.
\end{eqnarray}
To obtain an exact solution of Eq. (\ref{qbarra}),
let us consider the de Sitter spacetime where $a=e^{Ht}$, and 
\begin{equation}
\eta=-\frac{e^{-Ht}}{H}=\frac{1}{Ha(t)},\;\dot{a}=-\frac{1}{\eta},\;\ddot{a}=-\frac{H}{\eta}.
\end{equation}
Plugging these relations into Eq. (15) we obtain 
\begin{equation}
\bar{q}_{\lambda ik}''+\frac{1}{\eta}\bar{q}_{\lambda ik}'+\left\{ k^{2}-\frac{1}{\eta^{2}}\left[\frac{(D-1)^{2}}{4}-\frac{a^{2}\mu^{2}+12H^{2}\xi}{H^{2}}\right]\right\} q_{\lambda ik}=0.
\end{equation}
 which can be written as 
\begin{equation}
\left\{ \frac{d^{2}}{d(k\eta)^{2}}+\frac{1}{(k\eta)}\frac{d}{d(k\eta)}+\left[1-\frac{\nu^{2}}{(k\eta)^{2}}\right]\right\} \bar{q}_{\lambda ik}=0.\label{bessel}
\end{equation}
 where $R=12H^{2}$, and 
\begin{equation}
\nu^{2}=\frac{(D-1)^{2}}{4}-\frac{a^{2}\mu^{2}+12H^{2}\xi}{H^{2}}.
\end{equation}
Equation (\ref{bessel}) is a Bessel equation with solutions given
by $J_{\nu}(k|\eta|)$ and $N_{\nu}(k|\eta|)$. The two linearly independent
solutions for $q$ are: 
\begin{equation}
q_{\lambda ik}=\begin{cases}
a^{\frac{-(D-1)}{2}}J_{\nu}(k|\eta|)\\
a^{\frac{-(D-1)}{2}}N_{\nu}(k|\eta|)
\end{cases}.
\end{equation}
Finally, according to Ref.\cite{Finelli:1999dk,Bertoni:1997qb}, a
particular solution of Eq. (\ref{MPa}) reads {\small 
\begin{eqnarray}
\rho=a^{\frac{-(D-1)}{2}}\left[AJ_{\nu}^{2}+BN_{\nu}^{2}+\left(AB-\frac{\pi^{2}}{4H^{2}}\right){}^{\frac{1}{2}}J_{\nu}N_{\mu}\right]^{\frac{1}{2}},\label{solutionAB}
\end{eqnarray}
}where $A$ and $B$ are real constants. The fixing of these constants is related to the choice of our vacuum. This is due to the fact that the construction
of particle states and the choice of the vacuum is not unique in curved spaces as the one used here. This is important since the production
of particles can be inferred only after we choose some vacuum to compare with our physical solution. A natural choice is the Bunch-Davies vacuum, which is
the adiabatic vacuum at early times ($t\to -\infty$)\cite{Bertoni:1997qb}. For this adiabatic vacuum at early
times $A=B=\pi/2H$ and $\rho$ becomes 
\begin{eqnarray}
\rho=(H\eta)^{\frac{(D-1)}{2}}\sqrt{\frac{\pi}{2H}}\left[J_{\nu}^{2}+N_{\nu}^{2}\right]^{\frac{1}{2}},\label{solution}
\end{eqnarray}
for $a=1/H\eta$. This is the general solution for the scalar field
for arbitrary $D$. We can see that for $D=4$ our solution gives
\begin{eqnarray}
\rho=a^{-\frac{3}{2}}\sqrt{\frac{\pi}{2H}}\left[J_{\nu}^{2}+N_{\nu}^{2}\right]^{\frac{1}{2}},\label{solution}
\end{eqnarray}
 with 
\begin{equation}
\nu^{2}=\frac{9}{4}-\frac{a^{2}\mu^{2}+12H^{2}\xi}{H^{2}}.
\end{equation}
This result agrees with the solution found in Ref. \cite{Bertoni:1997qb} for $D=4$. 

\section{Concluding remarks}

In this paper we used the Lewis and Riesenfeld method to obtain the
time-dependent Schrödinger states emerging from the quantization of
the scalar field in the $D$-dimensional de Sitter spacetime. There
is a similarity between the equations found here and the ones for
the electromagnetic field. However, differently from the latter case,
we have a parameter $\xi$ that controls the conformality of the system. 
A general solution for arbitrary $D$ and $\xi$ is therefore very
usefull to analyze the physics of the problem.

Let us first analyze the $D=3$ case. This must becomes important for condensed matter systems. 
The solution found is identical to that one of the gauge field in $D=4$ \cite{Finelli:1999dk}
if we fix $\xi=\mu=0$. One could mistaken conclude that $\rho=constant$
is the only solution to the problem. However, we should remember that
Eq. (\ref{solution}) is obtained by considering a system evolving
to a vacuum state in the limit $t\to-\infty$ what surely is not the
case here. Therefore the solution for $D=3$ must be given by Eq.
(\ref{solutionAB}) and the constants $A,\, B$ must not be fixed
by fundamental arguments but from initial conditions in the referred
system.The next case is $D=4$, and we have seen that our results agree we 
the one in the literature \cite{Bertoni:1997qb}. Here we can seen clearly 
that the choice of $\xi$ controls the conformality of the system. If we
choose $\xi=1/6$ we obtain a conformal action and the trivial $\rho=constant$, as expected. 

A very intriguing consequence of the results obtained here is for
extra dimension physics. This has gained a lot of attention due superstring
theory \cite{Polchinski:1998rq,Polchinski:1998rr} and Randall Sundrum
models \cite{Randall:1999vf,Randall:1999ee}. In such model our universe
is conceived as a brane in a five dimensional space. If we choose a value for
$\xi$ to keep the conformal invariance in the brane($D=4$), we must loose the 
conformal invariance from the $D=5$ viewpoint. A de Sitter space
time would therefore imply a thermal bath for the comoving referentials
in this enlarged space. Therefore, at least in principle, this can
add an effective temperature in the membrane that could contributes
to the overall dynamics of the universe. However we should point
that from this viewpoint any field should contributes for this 
effective temperature and it is not clear for the authors
how to separate the contributions. 

At last, we would like to point out that the procedure described here
can be used to trace the present properties of the quantum scalar
field back to the recombination era in an arbitrary $D$-dimensional
universe. This would be a much more interesting phenomenological result.

\section{Acknowledgments}

The authors would like to thank Jailson S. Alcaniz for useful discussions.
We acknowledge the financial support provided by Fundação Cearense
de Apoio ao Desenvolvimento Cientí fico e Tecnológico (FUNCAP), the
Conselho Nacional de Desenvolvimento Cientí fico e Tecnológico (CNPq)
and FUNCAP/CNPq/PRONEX.


\begin{thebibliography}{References}
\bibitem{Polchinski:1998rq} J.~Polchinski, 
\textit{Cambridge, UK: Univ. Pr. (1998) 402 p}

\bibitem{Polchinski:1998rr} J.~Polchinski, 
\textit{Cambridge, UK: Univ. Pr. (1998) 531 p}


\bibitem{Berkovits:2000fe} N.~Berkovits, 
 JHEP \textbf{0004}, 018 (2000) {[}arXiv:hep-th/0001035{]}. 

\bibitem{Birrell:1982ix} N.~D.~Birrell and P.~C.~W.~Davies,
\textit{Cambridge, Uk: Univ. Pr. ( 1982) 340p}




\bibitem{Hawking:1974rv} S.~W.~Hawking, 
 Nature \textbf{248}, 30 (1974). 





\bibitem{Crispino:2007eb} L.~C.~B.~Crispino, A.~Higuchi and G.~E.~A.~Matsas,
 Rev.\ Mod.\ Phys.\ \textbf{80}, 787 (2008) {[}arXiv:0710.5373
{[}gr-qc{]}{]}. 





\bibitem{Saharian:2009ii} A.~A.~Saharian and T.~A.~Vardanyan,
 Class.\ Quant.\ Grav.\ \textbf{26}, 195004 (2009) {[}arXiv:0907.1149
{[}hep-th{]}{]}. 





\bibitem{Parker:1968mv} L.~Parker, 
 Phys.\ Rev.\ Lett.\ \textbf{21}, 562 (1968). 





\bibitem{Parker:1969au} L.~Parker, 
 Phys.\ Rev.\ \textbf{183}, 1057 (1969). 





\bibitem{Parker:1971pt} L.~Parker, 
 Phys.\ Rev.\ D \textbf{3}, 346 (1971) {[}Erratum-ibid.\ D \textbf{3},
2546 (1971){]}. 





\bibitem{Randall:1999vf} L.~Randall and R.~Sundrum, {}``An alternative
to compactification,'' Phys.\ Rev.\ Lett.\ \textbf{83}, 4690 (1999)
{[}arXiv:hep-th/9906064{]}. 





\bibitem{Randall:1999ee} L.~Randall and R.~Sundrum, {}``A large
mass hierarchy from a small extra dimension,'' Phys.\ Rev.\ Lett.\ \textbf{83},
3370 (1999) {[}arXiv:hep-ph/9905221{]}. 





\bibitem{Kehagias:2000au} A.~Kehagias and K.~Tamvakis, {}``Localized
gravitons, gauge bosons and chiral fermions in smooth spaces generated
by a bounce,'' Phys.\ Lett.\ B \textbf{504}, 38 (2001) {[}arXiv:hep-th/0010112{]}.





\bibitem{Bazeia:2008zx} D.~Bazeia, A.~R.~Gomes, L.~Losano and
R.~Menezes, {}``Braneworld Models of Scalar Fields with Generalized
Dynamics,'' Phys.\ Lett.\ B \textbf{671}, 402 (2009) {[}arXiv:0808.1815
{[}hep-th{]}{]}. 





\bibitem{Landim:2010pq} R.~R.~Landim, G.~Alencar, M.~O.~Tahim,
M.~A.~M.~Gomes and R.~N.~C.~Filho, {}``Dual Spaces of Resonance
In Thick $p-$Branes,'' arXiv:1010.1548 {[}hep-th{]}. 





\bibitem{Landim:2011ki} R.~R.~Landim, G.~Alencar, M.~O.~Tahim
and R.~N.~C.~Filho, 
 JHEP \textbf{1108}, 071 (2011) {[}arXiv:1105.5573 {[}hep-th{]}{]}.





\bibitem{MM} A. M. de M. Carvalho, C. Furtado, I. A. Pedrosa, Physical
Review D 70, 123523 (2004). 




\bibitem{Lewis:1968tm} H.~R.~Lewis and W.~B.~Riesenfeld, 
 J.\ Math.\ Phys.\ \textbf{10}, 1458 (1969). 




\bibitem{pedrosa1} C. E. F. Lopes, I. A. Pedrosa, C. Furtado and
A. M. de M. Carvalho, J. Math. Phys., 50, 083511 (2009).




\bibitem{Bertoni:1997qb} C.~Bertoni, F.~Finelli and G.~Venturi,
 Phys.\ Lett.\ A \textbf{237}, 331 (1998) {[}arXiv:gr-qc/9706061{]}.








\bibitem{pedrosa2} I. A. Pedrosa, Claudio Furtado and Alexandre Rosas,
Europhys. Lett. 94, 30002 (2011).







\bibitem{Milne} E. W. Milne, Phys. Rev. 35 (1930) 863.




\bibitem{Pinney} E. Pinney, Proc. Am. Math. Soc. 1 (1950) 681.




\bibitem{Alencar:2011nw} G.~Alencar, I.~Guedes, R.~R.~Landim
and R.~N.~C.~Filho, 
 arXiv:1107.2558 {[}hep-th{]}. 





\bibitem{Trivedi:2004ny} S.~P.~Trivedi, 
 Pramana \textbf{63}, 777 (2004). 





\bibitem{Carinena} J. F. Cariñena, J. de Lucas, Int. J. Geom. Methods
Mod. Phys., 6:4 (2009), 683\textendash{}699.




\bibitem{Pedrosa_97} I. A. Pedrosa, Phys. Rev. A 55, 3219 (1997).




\bibitem{Pedrosa_05} I. A. Pedrosa, A. Rosas and I. Guedes, J. Phys
Gen. 38, 7757 (2005).




\bibitem{Finelli:1999dk} F.~Finelli, A.~Gruppuso and G.~Venturi,
 Class.\ Quant.\ Grav.\ \textbf{16}, 3923 (1999) {[}arXiv:gr-qc/9909007{]}.

\end{thebibliography}
\end{document}